\documentstyle[12pt]{article}

\newcommand {\beq}{\begin{equation}}
\newcommand {\eeq}{\end{equation}}
\newcommand {\beqa}{\begin{eqnarray}}
\newcommand {\eeqa}{\end{eqnarray}}         %Equation version
\newcommand {\beqs}{\begin{eqnarray*}}
\newcommand {\eeqs}{\end{eqnarray*}}
\newcommand {\bds}{\begin{displaymath}}
\newcommand {\eds}{\end{displaymath}}
\newcommand {\n}{\nonumber\\}
\newcommand {\nn}{\nonumber}
\newcommand{\no}{\noindent}
\newcommand {\eqn}[1]{(\ref{#1})}

\newcommand {\bebb}{\begin{thebibliography}}
\newcommand {\eebb}{\end{thebibliography}}      %Reference version
\newcommand {\bbit}{\bibitem}
\def\al{\alpha}
\def\bt{\beta}

\def\dl{\delta}
\def\Dl{\Delta}

\def\eps{\epsilon}

\def\gm{\gamma}

\def\Lm{\Lambda}

\def\ph{\phi}
\def\Ph{\Phi}

\def\Ps{\Psi}

\def\sgm{\sigma}
\def\Sgm{\Sigma}

\def\tht{\theta}

\def\tl{\tilde}  %%%%%%%%%%For over letters%%%%%%%%%

\def\p{\partial}

\def\lrk{\left(}
\def\rrk{\right)}
\def\lsk{\left[}
\def\rsk{\right]}
\def\ot{\otimes}

\def\op{\oplus}

\def\journal#1&#2(#3){\unskip, \sl #1\ \bf #2 \rm(19#3) }
\def\andjournal#1&#2(#3){\sl #1~\bf #2 \rm (19#3) }

\def\npb#1#2#3{Nucl. Phys. {\bf B#1}, (#2) #3}

\def\plb#1#2#3{Phys. Lett. {\bf B#1}, (#2) #3}

\def\cmp#1#2#3{Commun. Math. Phys. {\bf #1}, (#2) #3}

\def\lmp#1#2#3{Lett. Math. Phys. {\bf #1}, (#2) #3}

\begin{document}

%\begin{titlepage}

\begin{flushright}
hep-th/0109009
\end{flushright}

\vskip 1cm

\begin{center}
%\title
{\Large\bf Twisted $sl(3, {\bf C})^{(2)}_k $ Current Algebra: Free Field\\
Representation and Screening Currents}

\vspace{1cm}

%\author{
{\normalsize\bf
Xiang-Mao Ding $^{a,b}$ %{\thanks{E-mail:xmding@maths.uq.edu.au}},
Mark D. Gould $^a$ and Yao-Zhong Zhang $^a$
}

{\em $^a$ Center of Mathematical Physics, Department of Mathematics, \\
University of Queensland, Brisbane, Qld 4072, Australia}
\\
{\em $^b$ Institute of Applied Mathematics, Academy of Mathematics 
and System Sciences; Chinese Academy of Sciences, P.O.Box 2734, 
100080, China.}

\end{center}

\date{}

%\maketitle

\vspace{2cm}

\begin{abstract}

Motivated by application of twisted current algebra in description of the 
entropy of $Ads_3$ black hole, we investigate the simplest twisted current 
algebra $sl(3, {\bf C})^{(2)}_k$. Free field representation of the twisted
algebra, and the corresponding twisted Sugawara energy-momentum tensor 
are obtained by using three $(\bt,\gm)$ pairs and two scalar fields. Primary 
fields and two screening currents of the first kind are presented.   

\end{abstract}

\vspace{1cm}

%%%%%PACS: 11.25Hf; 11.30.Rd; 03.65Fd; 02.20.Hj.

\vspace{0.5cm}

%\end{titlepage}

%\setcounter{section}{1}
%\setcounter{equation}{0}
\section{Introduction}

Virasoro algebra and affine algebras are algebraic structures in 
conformal field theories (CFT) in two dimensional spacetime~
\cite{BPZ,Kac,Ph}. They also play a central role in the study of string 
theory~\cite{Pol}. The free field realization is a common approach 
used in both conformal field theories 
and representation theory of affine Lie algebras~\cite{Wak}. 
The free field representations for untwisted affine algebra have been 
extensively studied. 
The simplest case $sl(2, {\bf C})^{(1)}$ was first treated 
in~\cite{Wak}, and generalization to $sl(n, {\bf C})^{(1)}$ was 
given in~\cite{BOo,FF1,FF2,FF3,BMP1,BMP2,BMP3,Ito,Kur,Fren,PRYu}. 
For twisted cases, however, little has been known. The recently 
study shows that twisted affine algebras are useful in the 
description of the entropy of $Ads_3$ black hole~\cite{SFF}. So it 
is desirable to further investigate the free field representations of 
twisted affine algebras. Let us remark that throughout this paper
we are dealing with algebras over the complex field.

In this letter we consider the simplest twisted affine algebra $sl(3,
{\bf C})^{(2)}$, and  construct the free field representation for 
this algebra. Moreever we also 
give the screening currents and primary fields. We remark that our result 
is different from the one obtained in~\cite{MS}.  

\section{Notation: twisted $sl(3, {\bf C})^{(2)}_k$ affine currents}

Let us start with some basic notations of twisted affine 
algebras~\cite{Kac}. 
Let $g$ be a simple finite-dimensional Lie algebra and $\sgm $ be an 
automorphism of $g$ satisfying $\sgm ^r =1$ for a positive integer $r$, 

then $g$ can be decomposed into the form: 

\beq
g=\oplus_{j\in {\bf Z}/r{\bf Z}}~~ g_j, 
\eeq

\no where $g_j$ is the eigenspace of $\sgm $ with eigenvalue 
$e^{2j{\pi} i/r}$, 
and $[g_i, g_j]\subset g_{(i+j)~mod~r}$~, then $r$ is called 
the order of the automorphism. 

For our purpose, we consider here only the simplest twisted affine algebra 
$A^{(2)} _2$, the $2$-order twisted affine algebra of 
$A_2=sl(3, {\bf C})$. Let 
${\tl \al }_1=\eps _1- \eps _2,~~{\tl \al}_2 =\eps _2 -\eps _3$ 
be the two simple roots of $sl(3, {\bf C})$ with normalization  
${\tl \al} ^2 _1={\tl \al} ^2 _2= 2,~~{\tl \al} _1 \cdot {\tl \al}_2 =-1$, 
and $\tht ^0= {\tl \al}_1 +{\tl \al }_2$. Set  
${\al }_1=\frac{1}{2}\left({\tl \al}_1 +{\tl \al}_2 \right)$ and 
${\al }_2=\frac{1}{2}\left( {\tl \al}_1 -{\tl \al}_2 \right)$. 
Then we have ${\al }_1 ^2=\frac{1}{2},~~{\al }_2 ^2=\frac{3}{2}$, 
and ${\al }_1\cdot {\al }_2=0$. We can write  

\beq
g=g_0\op g_1
\eeq

\no where $g_0$ is a fixed point subalgebra under the automorphism, 
while $g_1$ is a five dimensional representation of $g_0$, $g_0$ 
and $g_1$ satisfy $[g_i, g_j]\subset g_{(i+j)~mod ~2}$. 
Let ${\tl e}_{ij}$ are the matrix with entry $1$ at the $i$-th row 
and $j$-th column, and zero elsewhere. We choose the basis of $g_0$ as,  

\beqa
 e={\sqrt 2}({\tl e}_{12}+{\tl e}_{23})=
    {\sqrt 2}({\tl e}_1+{\tl e}_2); ~~~~~~ 
 f={\sqrt 2}({\tl e}_{21}+{\tl e}_{32})={\sqrt 2}({\tl f}_1+{\tl f}_2); \n
 h=2({\tl e}_{11}-{\tl e}_{33})=2( {\tl h}_1+{\tl h}_2).
\eeqa 

\no The basis for $g_1$ is taken to be

\beqa
 {\tl e}={\sqrt 2}( {\tl e}_{12} -{\tl e}_{23} )=
     {\sqrt 2}{\tl e}_1 -{\tl e}_2, ~~~~~~
 {\tl f}={\sqrt 2}({\tl e}_{21}-{\tl e}_{32})=
      {\sqrt 2}({\tl f}_1-{\tl f}_2), \n 
 {\tl E}=-2({\tl e}_1 {\tl e}_2 - {\tl e}_2 {\tl e}_1)=
     -2{\tl e}_{13}, ~~~~~~ 
 {\tl F}=-2({\tl f}_1 {\tl f}_2 -{\tl f}_2 {\tl f}_1)=-2{\tl e}_{31},\n
 {\tl h}=2({\tl e}_{11}-{\tl 2}e_{22}+{\tl e}_{33})
 =2({\tl h}_1-{\tl h}_2).
\eeqa

\no Then we have the following relations:

\beqa
&& [h,e]=2e; ~~~~~~  [h,f]=-2f;\n
&& [{\tilde e},{\tl F}]=2f; ~~~~~~
 [{\tl f},{\tl E}]=-2e; \n 
&& [{\tilde h},{\tl e}]=6e; ~~~~~~ 
 [{\tl h},{\tl f}]=-6f; \n
&& [e,f]=h=[{\tilde e},{\tl f}];~~~~~~[{\tl E},{\tl F}]=2h;\n 
&& [e,{\tl e}]=2{\tl E}; ~~~~~~[f,{\tl f}]=-2{\tl F}; \\
&& [f,{\tl E}]=2{\tl e};~~~~~~ [e,{\tl F}]=-2{\tl f} \n
&& [{\tl h},e]=6{\tl e}; ~~~~~~ [{\tl h},f]=-6{\tl f};\n
&& [h,{\tl e}]=2{\tl e}; ~~~~~~ [h, {\tl f}]=-2{\tl f}; \n 
&& [h,{\tl E}]=4{\tl E}; ~~~~~~ [h, {\tl F}]=-4{\tl F}; \n
&& [e,{\tl f}]={\tl h}=-[f, {\tl e}]. \nn
%&& [{\tilde e},{\tl E}]=0=[{\tl f},{\tl F}]; \n
%&& [{\tilde h},{\tl E}]=0=[{\tl h},{\tl F}]; \n
%&& [e,{\tl E}]=0=[f,{\tl F}]; \n
%&& [h,{\tl h}]=0.
\eeqa

\no All other commutators are zero. It is easy to verify that 
the following operator is the quadratic Casimir of $sl(3, {\bf C})$ in 
the above basis

\beq
C_2=h^2 +\frac{1}{3}{\tl h} ^2 + 4fe +4 {\tl f}{\tl e}+4{\tl E}{\tl F}.
\eeq

\no This quadratic Casimir element is useful in sequel to construct the 
twisted energy-momentum stress tensor. 

The commutators of $sl(3, {\bf C})^{(2)}_k$ can be expressed as

\beq
\lsk z^m\ot X, z^n \ot Y \rsk=z^{m+n}\ot \lsk X,Y \rsk 
+2km \dl _{m+n,0}\frac{(X|Y)}{2}.
\label{TKM}
\eeq 
 
\no Denote the currents corresponding to 
$e,~h,~f$ by $j^+(z),~j^0(z),~j^-(z)$, and to 
${\tl e},~{\tl h},~{\tl f},~{\tl E},~{\tl F}$ by $J^+(z),~J^0(z),~J^-(z),
~J^{++}(z),~J^{--}(z)$, respectively. Then \eqn{TKM} can be written in terms 
of the following OPE's:

\beqa
j^+ (z) j^-(w)=&&\frac{4k}{(z-w)^2}+\frac{1}{(z-w)}j^0 (w)+\ldots;\n
j^0 (z) j^{\pm} (w)=&&\frac{\pm 2}{(z-w)}j^{\pm} (w)+\ldots;\n
j^0 (z) j^{0}(w)=&&\frac{8k}{(z-w)^2}+\ldots;\n
J^+ (z) J^-(w)=&&\frac{4k}{(z-w)^2}+\frac{1}{(z-w)}j^0 (w)+\ldots;\n
J^{++} (z) J^{--}(w)=&&\frac{4k}{(z-w)^2}+\frac{2}{(z-w)}j^0 (w)+\ldots;\n
J^0 (z) J^{\pm} (w)=&&\frac{\pm 6}{(z-w)}j^{\pm} (w)+\ldots;\n
J^0 (z) j^{\pm} (w)=&&\frac{\pm 6}{(z-w)}J^{\pm} (w)+\ldots;\n
J^+ (z) J^{--}(w)=&&\frac{2}{(z-w)}j^- (w)+\ldots;\n
J^- (z) J^{++}(w)=&&\frac{-2}{(z-w)}j^+ (w)+\ldots; \n
j^+ (z) J^+(w)=&&\frac{2}{(z-w)}J^{++} (w)+\ldots; \n
j^- (z) J^-(w)=&&\frac{-2}{(z-w)}J^{--} (w)+\ldots;\n
j^+ (z) J^-(w)=&&\frac{1}{(z-w)}J^{0} (w)+\ldots;\n
J^+ (z) j^-(w)=&&\frac{1}{(z-w)}J^{0} (w)+\ldots;\n
j^+ (z) J^{--}(w)=&&\frac{-2}{(z-w)}J^{-} (w)+\ldots;\n
j^- (z) J^{++}(w)=&&\frac{2}{(z-w)}J^{+} (w)+\ldots;\n
j^0 (z) J^{\pm} (w)=&& \frac{\pm 2}{(z-w)}J^{\pm } (w)+\ldots;\n
j^0 (z) J^{\pm \pm} (w)=&& \frac{\pm 4}{(z-w)}J^{\pm \pm} (w)+\ldots;\n
J^0 (z) J^{0}(w)=&& \frac{24k}{(z-w)^2}+\ldots. \label{ope}
\eeqa

\no All other OPE's contain trivial regular terms only. 
Here and throughout $"\ldots"$ stands for regular terms.

%J^0 (z) J^{\pm \pm} (w)=&&+\ldots;\n
%J^{+} (z) J^{++} (w)=&&+\ldots; \n
%J^- (z) J^{--} (w)=&&+\ldots; \n
%j^+ (z) J^{++}(w)=&&+\ldots; \n
%j^- (z) J^{--}(w)= &&+\ldots;\n
%j^0 (z) J^{0} (w)=&& +\ldots . \nn

%\setcounter{section}{3}
%\setcounter{equation}{0}
\section{Wakimoto realization of the twisted affine currents}

To obtain a free field realization of the twisted $sl(3, {\bf C})^{(2)}_k$ 
currents, we first construct a Fock space of $sl(3, {\bf C})$ in 
the basis given in 
section $2$. The Fock space is constructed by the repeated actions of 
$f,~ {\tl f},~{\tl F}$ on the highest weight state $v_{\Lm}$, which is 
determined by  
\beqa
&&ev_{\Lm}={\tl e}v_{\Lm}={\tl E}v_{\Lm}=0;\n
&&hv_{\Lm}=(\Lm, \al _1)v_{\Lm}, \hskip 0.5cm 
  {\tl h}v_{\Lm}=(\Lm, \al _2)v_{\Lm}.
\eeqa  
Set $|l,m,n>=f^l{\tl F}^m {\tl f}^n v_{\Lm}$. We find
\beqa
f|l,m,n>=&&|l+1,m,n>; \n
{\tl f}|l,m,n>=&&2l|l-1,m+1,n>+|l,m,n+1>; \n
{\tl F}|l,m,n>=&&|l,m+1,n>;\\
h|l,m,n>=&&-\left[ 2l+4m+2n-(\Lm , \al _1) \right]|l,m,n>; \n
{\tl h}|l,m,n>=&&-6l|l-1,m,n+1>-6l(l-1)|l-2,m+1,n> \n
           &&-6n|l+1,m,n-1>-6n(n-1)|l,m+1,n-2> \n 
	   &&+(\Lm,\al _2) |l,m,n>\nn
\eeqa

\no For other generators, we obtain 

\beqa
e|l,m,n>=&&-l\left[(l-1)+4m+2n-(\Lm ,\al _1)\right] |l-1,m,n> \n
               &&-2m|l,m-1,n+1>-3n(n-1)|l+1,m,n-2> \n
	       &&-4n(n-1)(n-2)|l,m+1,n-3> \n
	       &&+n(\Lm, \al _2) |l,m,n-1>; \n
{\tl e}|l,m,n>=&&-n\left[ 6l+(n-1) -(\Lm, \al _1) \right]|l,m,n-1> \n
               &&+2m|l+1,m-1,n>-6ln(n-1)|l-1,m+1,n-2>; \n
               &&-2l(l-1)(l-2)|l-3,m+1,n>+l(\Lm,\al _2)|l-1,m,n> \n
               &&-3l(l-1)|l-2,m,n+1>;  \\
{\tl E}|l,m,n>=&&-2m\left[2l+2n+2(m-1) -m (\Lm, \al _1)\right]|l,m-1,n> \n
               &&+2ln\left[(n-1)+3(l-1)-(\Lm, \al _1)\right] |l-1,m,n-1> \n
	       &&+6l(l-1)n(n-1)|l-2,m+1,n-2> \n
	       &&+2l(l-1)(l-2)|l-3,m,n+1> \n
	       &&+l(l-1)(l-2)(l-3)|l-4,m+1,n> \n
	       &&-2m(m-1)(m-2)|l+1,m,n-3> \n
	       &&-3m(m-1)(m-2)(m-3)|l,m+1,n-4> \n
	       &&-l(l-1)(\Lm, \al _2)|l-2,m,n> \n
	       &&+m(m-1)(\Lm, \al _2)|l,m,n-2>.\nn
\eeqa

Now we use the above representation as a tool to construct, 
via the induced procedure \cite{BMP3}, free field representations of
the twisted $sl(3, {\bf C})^{(2)}_k$ current algebra. 
In the Fock space, if we regard 
$\gm _0, ~\gm _1,~\gm _2$ as $f,~{\tl f},~{\tl F}$, 
and $\bt _0, ~\bt _1,~\bt _2$ as 
$-\frac{\p}{\p f}$, $-\frac{\p}{\p {\tl f}}$,$-\frac{\p}{\p {\tl F}}$, 
respectively, then we obtain a realization of 
the non-affine algebra in terms of the differential operators

\beqa
e=&&-\gm _0 \lrk \bt ^2 _0 +3 \bt ^2 _1  \rrk
    - 2 \gm _1 \lrk \bt _0 \bt _1 -\bt _2\rrk \n
  &&- 4 \gm _2 \lrk \bt _0\bt _2 -\bt ^3 _1 \rrk 
  - (\Lm, \al _1) \bt  _0  - (\Lm, \al _2) \bt _1;\n
h=&&2 \bt _0 \gm _0 + 2 \bt _1 \gm _1 
    + 4 \bt _2 \gm _2 +(\Lm, \al _1);\n
f=&&\gm _0 (z);\n
{\tl E} =&&2\gm _0 \lrk \bt ^3 _1 - 2\bt _0 \bt _2
    -3\bt ^2 _0 \bt _1 \rrk \n
         &&- 2\gm _1 \lrk \bt ^3 _0  +\bt _0 \bt ^2 _1 
    +2\bt _1\bt _2  \rrk \n
         &&+\gm _2 \lrk \bt ^4 _0  -3\bt ^4 _1
    +6\bt ^2 _0 \bt ^2 _1 -4\bt ^2 _2 \rrk \n
         &&-2(\Lm, \al _1) \lrk \bt _0 \bt _1  +\bt _2  \rrk 
	   -(\Lm, \al _2) \lrk \bt ^2 _0  - \bt ^2 _1 \rrk;  \n
{\tl e}=&&-2\gm _0 \lrk 3\bt _0 \bt _1+\bt _2  \rrk
          - \gm _1 \lrk 3\bt ^2_0 +\bt ^2 _1\rrk \n
        &&+2 \gm _2 \lrk \bt ^3 _0+3\bt _0 \bt ^2 _1 \rrk 
          -(\Lm, \al _1) \bt _1 (z) -(\Lm, \al _2) \bt _0 ;\n
{\tl h}=&&6 \gm _0 \bt _1+ 6 \gm _1 \bt _0
        - 6 \gm _2 \lrk \bt ^2 _0 + \bt ^2 _1\rrk +(\Lm, \al _2);\n
{\tl f}=&&\gm _1 - 2 \gm _2 \bt _0; \n
{\tl F}=&&\gm _2;\nn
\eeqa
 
It turns out that to get a free field realization of the twisted currents, 
we have to interchange the $\bt _i$ with $\gm _i$ in the above expressions, 
and at the same time interchange $e$ with $f$. 
Now we introduce three $\bt \gm$  
pairs and two scalar fields $\ph _a$, $a=1,~2$. $(\bt _i;~\gm _i)$ pairs 
have conformal dimension $(1;~0)$. 

\beqa
\bt _i (z)\gm _j(w)=-\gm _j(z) \bt _i(w)=-\frac{\dl _{ij}}{z-w},  
\hskip 0.5cm i,j=0,1,2 \n
\ph _a (z) \ph _b (w)=-2 \dl _{a,b} ln(z-w), \hskip 0.5cm a,b=0,1
\eeqa

\no Introduce the notation $\vec{e}_1 =\frac{1}{2}(1,1)$; 
$\vec{e}_2=\frac{\sqrt{3}}{2}(1,-1)$; and 
$\vec{\Ph}=(\ph _0, \ph _1)$. Then we have 

\beqs
  \vec{e}_1\cdot \vec{\Ph} (z) \vec{e}_1\cdot \vec{\Ph} (w)=-ln(z-w); 
  \hskip 0.5cm \vec{e}_2\cdot \vec{\Ph} (z) 
     \vec{e}_2\cdot \vec{\Ph} (w)=-3ln(z-w);\\
  \vec{e}_1\cdot \vec{\Ph} (z) \vec{e}_2\cdot \vec{\Ph} (w)=0.
\eeqs

\no With the help of the differential operator representation of 
the non-affine algebra, we find the Wakimoto realization of 
$sl(3, {\bf C})^{(2)}_k$ in terms of the eight free fields:  

\beqa
j^+(z)=&&\bt _0 (z);\n
j^0(z)=&&2 \bt _0(z) \gm _0 (z)+ 2 \bt _1(z) \gm _1 (z)
+ 4 \bt _2(z) \gm _2 (z)
+ \frac{1}{\al _+}(\vec{e}_1 \cdot i\p \vec{\Ph} (z));\n
j^-=&&-\bt _0(z) \lrk \gm ^2 _0 (z)+3 \gm ^2 _1 (z) \rrk
- 2 \bt _1(z) \lrk \gm _0 (z)\gm _1 (z)-\gm _2(z)\rrk \n
&&- 4 \bt _2(z) \lrk \gm _0(z)\gm _2 (z)-\gm ^3 _1 (z)\rrk 
-4k \p \gm _0 (z) \n
&&- \frac{1}{\al _+}(\vec{e}_1 \cdot i\p \vec{\Ph} (z)) \gm _0 (z) 
- \frac{1}{\al _+}(\vec{e}_2 \cdot i\p \vec{\Ph} (z)) \gm _1 (z);\n
J^{++}(z)=&&\bt _2 (z); \\
J^+(z)=&&\bt _1 (z)- 2 \bt _2 (z)\gm _0 (z); \n
J^0(z)=&&6 \bt _0(z) \gm _1 (z)+ 6 \bt _1(z) \gm _0 (z)
- 6 \bt _2(z) \lrk \gm ^2 _0 (z)+ \gm ^2 _1 (z)\rrk \n
&&+ \frac{1}{\al _+}(\vec{e}_2 \cdot i\p \vec{\Ph} (z));\n
J^-=&&-2\bt _0(z) \lrk 3\gm _0 (z)\gm _1(z)+\gm _2 (z) \rrk
- \bt _1(z) \lrk 3\gm ^2_0 (z)+\gm ^2 _1(z)\rrk \n
&&+2 \bt _2(z) \lrk \gm ^3 _0(z)+3\gm _0 (z)\gm ^2 _1 (z) \rrk
 -4(k+1) \p \gm _1 (z)\n
&&- \frac{1}{\al _+}(\vec{e}_1 \cdot i\p \vec{\Ph} (z)) \gm _1 (z) 
- \frac{1}{\al _+}(\vec{e}_2 \cdot i\p \vec{\Ph} (z)) \gm _0 (z) ;\n
J^{--}=&&2\bt _0(z) \lrk \gm ^3 _1 (z)- 2\gm _0 (z)\gm _2(z)
-3\gm ^2 _0 (z)\gm _1 (z) \rrk \n
&&- 2\bt _1(z) \lrk \gm ^3 _0 (z) +\gm _0 (z)\gm ^2 _1 (z)
+2\gm _1(z)\gm _2 (z) \rrk \n
&&+\bt _2(z) \lrk \gm ^4 _0 (z) -3\gm ^4 _1(z)+6\gm ^2 _0 (z)\gm ^2 _1 (z)
-4\gm ^2 _2 (z) \rrk \n
&&- \frac{2}{\al _+}(\vec{e}_1 \cdot i\p \vec{\Ph} (z)) \lrk 
  \gm _0 (z)\gm _1 (z) +\gm _2 (z) \rrk \n
&&- \frac{1}{\al _+}(\vec{e}_2 \cdot i\p \vec{\Ph} (z)) \lrk 
  \gm ^2 _0 (z) - \gm ^2 _1(z) \rrk  \n
&&-8(k+1) \gm _0 (z)\p \gm _1 (z) -4k \p \gm _2 (z).\nn
\eeqa

\no Here $\al _+ =1/ \sqrt{8k+24}$, and normal ordering is 
implied in the expressions. It is straightforward to check that the above 
currents satisfy the OPEs given in last section. 

Note that the twisted currents have the following mode expansions: 

\beq
j^a (z)=\Sgm _{n \in Z} j^a _n z^{-n-1}; ~~~~~~ 
J^a (z)= \Sgm _{n \in Z+1/2} J^a _n z^{-n-1}. 
\eeq.

\no From the expressions of the currents, we find that 
the following relations hold 

\beqa
&&J^-(z)=\frac{1}{2}\frac{\p}{\p \gm _0(z)}J^{--}(z); \hskip 0.5cm 
  J^0(z)=-\frac{1}{3}\frac{\p}{\p \gm _0(z)}J^{-}(z);\n
&&J^+(z)=\frac{1}{2}\frac{\p}{\p \gm _0(z)}J^{0}(z); \hskip 0.5cm
  J^{++}(z)=-\frac{1}{2}\frac{\p}{\p \gm _0(z)}J^{+}(z); \\
&&j^0(z)=-\frac{\p}{\p \gm _0(z)}j^{-}(z); \hskip 0.5cm  
j^+(z)=\frac{1}{2}\frac{\p}{\p \gm _0(z)}j^{0}(z). \nn
\eeqa

Now let's examine the condition for our representation to be unitary.
We introduce the conjugate operation on the modes of the currents: 
\beq
T^{a \dagger}_n =T^{-a}_{-n}, \hskip 0.5cm k^{\dagger}=k, 
\eeq
where $T^a$ stands for $j^0, j^\pm, J^0, J^\pm$ or $J^{\pm\pm}$ and
$T^{-a}$ for $j^0, j^\mp, J^0, J^\mp$ or $J^{\mp\mp}$, respectively. From 
the OPEs \eqn{ope} we know that 
\beq
T^{-a}_{1/b},\hskip 0.3cm T^{a} _{-1/b}, \hskip 0.3cm 4k-bj^0 _0,
\eeq
where $a=+$ or $++$ and $-a=-$ or $--$, 
$b=1$ for $T^\pm=j^\pm$ and $b=2$ for $T^\pm=J^\pm$
or $J^{\pm\pm}$. Obviously, for a given $a$ and $b$,
$T^{-a}_{1/b},\,T^a_{-1/b}$ and $4k-bj^0_0$ form 
the $su(2)$ subalgebra of $sl(3, {\bf C}^{(2)}_k)$.
Firstly, the egenvalues of $4k-bj^0 _0$ must be integral for an unitary 
representation. Applying this to the state $v_\Lambda$, one obtains
\beq
j^0 _0\,v_\Lambda=(\Lambda,\al_1)\,v_\Lambda,
\eeq
which implies that
\beq
4k-b\,(\Lambda, \al_1) \in Z.
\eeq
As $v_\Lambda$ is a vacuum state, i.e. 
\beq
T_{1/b} ^{-a}v_\Lambda=0,
\eeq
where $a=+,\,++$,  thus we have the norm
\beqa
(T^a _{-1/b}v_\Lambda, T^a_{-1/b}v_\Lambda)&=&(v_\Lambda, 
     T^{-a}_{1/b}T^a _{-1/b}v_\Lambda)
    =(v_\Lambda, [T^{-a}_{1/b},T^a _{-1/b}]v_\Lambda)\n
&=&\left[4k-b\,(\Lambda, \al_1)\right](v_\Lambda, v_\Lambda).
\eeqa
Hence $4k\geq b (\Lambda, \al_1)$ with $b=1$ or $2$ corresponding to $a=+$
or $a=++$, respectively. We take $\Lambda$ to be dominant. Then 
$4k\geq 2(\Lambda, \al_1)\geq 0$. So the condition for our
representation to be unitary is $4k\in Z$ and $4k\geq 2(\Lambda,
\al_1)\geq 0$.

\section{Twisted stress energy tensor}

It is well known that Virasoro algebras are related to currents algebras via 
the so called Sugawara construction. In the present case, the 
twisted Sugawara construction of the energy-momentum tensor is given by 

\beqa
T(z)=&& \frac{1}{8(k+3)}:\left[ \frac{1}{2}j^0(z)j^0(z) 
  + \frac{1}{6}J^0(z)J^0(z) 
+2j^-(z)j^+(z) \right.\n
&& ~~~~~~~~~~~~~~~~\left.+ 2J^-(z)J^+(z) + 2J^{--}(z)J^{++}(z) \right]:,
\eeqa  

\no in which $:~:$ implies the normal ordering. The above expression can 
be rephrased through the $\bt \gm$ pairs and the scalar field $\vec{\Ph}$. 
We obtain 

\beqa
T(z)=&& - :\left[ \bt _0(z) \p \gm _0 (z)+ \bt _1(z) \p \gm _1 (z)+  
\bt _2(z) \p \gm _2 (z) \right]: \n
&&+ \frac{1}{2} :\left(  \vec{e}_1 \cdot i\p \vec{\Ph} (z) \right) ^2:
+\frac{1}{6}:\left( \vec{e}_2 \cdot i\p \vec{\Ph} (z) \right) ^2 : \n
&&-4\al _+ \left( \vec{e}_1 \cdot i {\p} ^2 \vec{ \Ph} (z) \right). 
\eeqa

\no Following the standard practice, 
we get the OPE of the energy-momentum tensor,  

\beq
T(z)T(w)=\frac{c/2}{(z-w)^4}+\frac{2T(w)}{(z-w)^2} 
  + \frac{\p T(w)}{z-w}+\ldots,
\eeq

\no where $c=8k/(k+3)$ is the central charge for the Virasoro algebra. 

\section{Twisted screening currents}
 
An important object in the free field approach is the screening current. 
Screening currents are a primary fields with conformal dimension $1$, 
and their integration give the screening charges. 
They commute with the affine currents up to a total derivative. 
These properties ensure that screening charges 
may be inserted into correlators while the conformal or 
affine ward identities remain intact. For the present case, 
we find the following screening currents
 
\beq
 S_{\pm}(z)=:\left[2\bt _2(z) \gm _1(z) -\bt _0 (z) \pm 
   \bt _1(z)\right] {\tl S}_{\pm}(z):, 
\eeq  
  
\no where

\beq 
{\tl S}_{\pm} (z)=e^{-2\al _+\vec{e}_1 \cdot i\vec{\Ph} (z) 
  \pm 2\al _+ \vec{e}_2 \cdot i\vec{\Ph} (z)}.
\eeq
 
The OPE of the twisted screening currents with the twisted affine 
currents are

\beqa
T(z)S_{\pm} (w)=\p _w \left(\frac{1}{z-w}S_{\pm}(w)\right)+\ldots;\n
j^+(z)S_{\pm}(w)=\ldots; ~~~~~~ j^0(z)S_{\pm}(w)=\ldots; \n
j^-(z)S_{\pm}(w)=\p _w \left(-\frac{1}{2\al ^2 _+}
\frac{1}{z-w}{\tl S}_{\pm}(w)\right)+\ldots;\n
J^+(z)S_{\pm}(w)=\ldots; ~~~~~~ J^0(z)S_{\pm} (w)=\ldots; \n
J^-(z)S_{\pm} (w)=\p _w \left(\pm \frac{1}{2\al ^2 _+}
\frac{1}{z-w}{\tl S}_{\pm}(w)\right)+\ldots;\\
J^{++}(z)S_{\pm}(w)=\ldots;\n
J^{--}(z)S_{\pm}(w)=\p _w \left(\pm \frac{1}{\al ^2 _+}\frac{1}{z-w}
\left[\gm _0(w) \pm \gm _1 (w)\right]{\tl S}_{\pm}(w)\right)+\ldots;\nn
\eeqa

\no The screening currents obtained here are the twisted versions of the 
first kind screening currents~\cite{BOo}.

\section{Twisted primary fields}

Primary fields are fundamental objects in conformal field theories. 
A primary field $\Ps$ has the following OPE with the energy-momentum 
tensor:    

\beq
T(z) \Ps (w)=\frac{h_{\Ps}}{(z-w)^2}\Ps (w)
    +\frac{\p _w \Ps (w)}{z-w}+\ldots,
\eeq

\no where the $h_{\Ps}$ is the conformal dimension of $\Ps$. Moreover 
the most singular part in the OPE of $\Ps$ with the affine currents 
only has a single pole. A special kind of the primary fields is highest 
weight state. In the present case, a highest weight state is one that is
annihilated by currents $j^+(z)$, $J^+(z)$ and $J^{++}(z)$. 
So the highest state has the form

\beq
V(z)=e^{a\vec{e}_1\cdot i\vec{\Ph} (z) + b\vec{e}_2 \cdot i\vec{\Ph} (z)}.
\eeq

\no It has conformal dimension $\Dl (V_{a;b})=\frac{1}{2}a(a+8 \al _+) 
+\frac{3}{2}b^2 $. Obviously $\Dl (V_{a,b})=\Dl (V_{a;-b})$. We find 
that only when  

\beqa
V(z)=V_{j,\pm}(z)=&&e^{\al _+ \left(\frac{j}{\al ^2 _1} 
   \vec{e}_1 \cdot i\vec{\Ph} (z) \pm 
\frac{j}{\al ^2 _2}\vec{e}_2 \cdot i\vec{\Ph} (z)\right)} \n
=&& e^{\al _+ \left(2j \vec{e}_1 \cdot i\vec{\Ph} (z) \pm 
\frac{2}{3} j\vec{e}_2 \cdot i\vec{\Ph} (z)\right)},
\label{hst}
\eeqa 
the number of fields produced by the repeated action of $j^-,~J^-,~J^{--}$ 
currents on this highest state are finite. 
The conformal dimensions corresponding 
to \eqn{hst} are 

\beq
\Dl _j =\frac{8}{3}j(j+3)\al ^2 _+=\frac{j(j+3)}{3(k+3)}.
\eeq

\no From this highest weight state, we obtain the following fields,

\beqa
\Ph ^m _{j,\pm} (z)=&&:\lsk \lrk -(\gm _0(z) \pm \gm _1 (z) \rrk ^{j-m} 
+ \Sgm_{n=1}(-1)^{j-m-n}\lsk (j-\frac{1}{2})(j-\frac{3}{2})\cdots  
  \right. \right. \n
&&\left.(j-\frac{(2n-1)}{2})\rsk ^{-1} \times \frac{m(m-1)
\cdots (m-2n+1)}{2\times 4 \times \cdots \times 2n} \n
&&\left. \times (\gm _0 (z)\pm \gm _1 (z) )^{j-m-2n}
(\gm ^2 _1(z)  \mp \gm _2 (z))^n \rsk V_{j;\pm}(z):; 
   \hskip 0.2cm m-2n \geq 0.
\eeqa

\no They are primary fields in the sense that their OPE with the 
energy-momentum tensor is
 
\beq
T(z) \Ph ^m _{j; \pm} (w)=\frac{\Dl _j}{(z-w)^2}\Ph ^m _{j; \pm} (w)
+\frac{\p _w \Ph ^m _{j; \pm}(w)}{z-w}+\ldots,
\eeq

\no and the OPE's with the fixed point subalgebra of $sl(3)^{(2)} _k$ are 

\beqa
j^+(z)\Ph ^m _{j;\pm} (w)&=&\frac {(j-m)}{z-w}\Ph ^{m+1} _{j;\pm} (w)
  +\ldots; \n
j^0(z)\Ph ^m _{j;\pm} (w)&=&\frac {2m}{z-w}\Ph ^{m} _{j;\pm} (w)+\ldots; \\
j^-(z)\Ph ^m _{j;\pm} (w)&=&\frac {(j+m)}{z-w}\Ph ^{m-1} _{j;\pm} (w)
  +\ldots. \nn
\label{jph}
\eeqa

\no From the expression of $\Ph ^m _{j;\pm} (z)$, we see that $j$ can only 
take integer values. The OPE's of $\Ph ^m _{j;\pm}$ with other currents are 
much more involved:

\beqa
J^+(z)\Ph ^m _{j;\pm} (w)=&&\frac{1}{z-w}\left( \pm (j-m) 
  \Ph ^{m+1} _{j;\pm} (w)
\mp \frac{(j-m)(j-m-1)}{j-1/2} \frac{\Ph ^{j-1} _{j;\pm} (w)}
   {\Ph ^j _{j;\pm} (w)}
\Ph ^{m+1} _{j-1;\pm} (w)\right)+\ldots; \n
J^-(z)\Ph ^m _{j;\pm} (w)=&&\frac{1}{z-w}
\left( \pm 2j \frac{\Ph ^{j-1} _{j;\pm} (w)}{\Ph ^j _{j;\pm} (w)}
\Ph ^{m} _{j;\pm} (w) 
\mp 2(j-m)(j-1/2) \frac{\Ph ^{j-2} _{j;\pm} (w)}{\Ph ^j _{j;\pm} (w)}
\Ph ^{m+1} _{j;\pm} (w) \right. \n
&&\hskip 1.5cm \pm 2(j-m)(j-2) 
\left(\frac{\Ph ^{j-1} _{j;\pm} (w)}{\Ph ^j _{j;\pm} (w)}\right)^2 
\Ph ^{m+1} _{j;\pm} (w) \n
&&\hskip 1.5cm \pm \left.\frac{(j-m)(j-m-1)}{j-1/2} 
\left(\frac{\Ph ^{j-1} _{j;\pm} (w)}{\Ph ^j _{j;\pm}(w)} \right)^3
\Ph ^{m+1} _{j-1;\pm} (w)\right)+\ldots; \n
J^0(z)\Ph ^m _{j;\pm} (w)=&&\frac{1}{z-w}
\left(  \pm 2j \Ph ^{m} _{j;\pm} (w) 
\mp 6(j-m) \frac{\Ph ^{j-1} _{j;\pm} (w)}{\Ph ^j _{j;\pm} (w)}
\Ph ^{m+1} _{j;\pm} (w) \right. \n
&&\hskip 1.5cm \pm \left.\frac{3(j-m)(j-m-1)}{j-1/2} 
\left(\frac{\Ph ^{j-1} _{j;\pm} (w)}{\Ph ^j _{j;\pm}(w)} \right)^2
\Ph ^{m+1} _{j-1;\pm} (w)\right)+\ldots; \\
J^{++}(z)\Ph ^m _{j;\pm} (w)=&&\frac{\mp 1}{z-w}
\frac{(j-m)(j-m-1)}{2(j-1/2)} 
\Ph ^{m+1} _{j-1;\pm} (w)+\ldots; \n
J^{--}(z)\Ph ^m _{j;\pm} (w)=&&\frac{1}{z-w}
\left( \pm 4(j-1)m  
\left( \frac{\Ph ^{j-1} _{j;\pm} (w)}{\Ph ^j _{j;\pm} (w)}\right)^2
\Ph ^{m} _{j;\pm} (w) \right.\n
&&\hskip 1.5cm \mp 4(j-1/2)m 
\left( \frac{\Ph ^{j-2} _{j;\pm} (w)}{\Ph ^j _{j;\pm} (w)} \right)
\Ph ^{m} _{j;\pm} (w) \n
&&\hskip 1.5cm\mp \frac{(j-m)(j-m-1)}{2(j-1/2)} 
\left(\frac{\Ph ^{j-1} _{j;\pm} (w)}{\Ph ^j _{j;\pm} (w)}\right)^4 
\Ph ^{m+1} _{j-1;\pm} (w) \n
&&\hskip 1.5cm \pm 2(j-m)(j-m-1)(j-1/2) \n
&&\hskip 1.5cm\times 
\left.\left[\left(\frac{\Ph ^{j-2} _{j;\pm} (w)}{\Ph ^j _{j;\pm}(w)} \right) 
-\left(\frac{\Ph ^{j-1} _{j;\pm} (w)}{\Ph ^j _{j;\pm}(w)} \right)^2\right]^2
\Ph ^{m+1} _{j-1;\pm} (w)\right)+\ldots.\nn
\eeqa
 
\vskip 1cm

\vskip 1cm

\no {\bf Acknowledgments:}

This work is financially supported by Australian Research Council. 
One of the authors (Ding) is also supported partly by the 
Natural Science Foundations of China and a Fund from AMSS. We thank
the anonymous refree for suggestions on the improvement of the
paper.

\bebb{99}

\bibitem{BPZ}%{1}
A.A. Belavin. A. M. Polyakov, A. B. Zamolodchikov, 
{\em Nucl. Phys.} {\bf B241}, (1984)333.

\bibitem{Kac}%{2}
V. G. Kac, {\it Infinite Dimensional Lie Algebras}, third ed.,
Cambridge University press, Cambridge 1990.

\bbit{Ph}%{3}
Ph.Di Francesco, P. Mathieu and D. Senechal, {\it Conformal Field Theory}, 
Springer, 1997.

\bbit{Pol}
J. Polchinski, {\it String Theory}, Cambridge University press, 
Cambridge 1998.

\bibitem{Wak}%{}
M. Wakimoto, {\it Commun. Math. Phys. {\bf 104}}, (1986)605.

\bibitem{BOo}
M.Bershsdsky and H.Ooguri, {\it Commun. Math. Phys.} {\bf126},(1989)49.

\bbit{FF1}
B. Feigin and E. Frenkel, {\it Russ. Math. Surv.} {\bf 43}:5 (1988)221. 

\bbit{FF2}
B. Feigin and E. Frenkel, \lmp {19}{1990}{307}.

\bbit{FF3}
 B. Feigin and E. Frenkel, \cmp {128}{1990}{161}. 

\bbit{BMP1}
P. Bouwknegt, J. McCarthy and K. Pilch, \plb {234}{1990}{297}.

\bbit{BMP2}
P. Bouwknegt, J. McCarthy and K. Pilch, \cmp {131}{1990}{125}.

\bbit{BMP3}
P. Bouwknegt, J. McCarthy and K. Pilch, {\it Prog. Phys. Suppl.} 
{\bf 102}, (1990) 67.

\bbit{Ito}
K. Ito, \plb {252}{1990}{69}

\bbit{Kur}
G. Kuroki, \cmp {142}{1991}{511}.

\bbit{Fren}
E. Frenkel, {\it Free Field Realizations in Representation Theory and 
Conformal Field Theory}, heh-th/9408109.

\bbit{PRYu}
J. L. Petersen, J. Rasmussen and M. Yu, \npb{502}{1997}{649}.

\bbit{SFF}
S. Fernando and F.Mansouri,\plb {505}{2001}{206}; hep-th/0010153.

\bbit{MS}
M. Szczesny, {\it Wakimoto modules for twisted affine Lie algebras}, 
math-QA/0106061.

\eebb

\end{document}